\documentclass[sigconf]{acmart}

\usepackage{multirow}

\AtBeginDocument{%
  \providecommand\BibTeX{{%
    \normalfont B\kern-0.5em{\scshape i\kern-0.25em b}\kern-0.8em\TeX}}}

\setcopyright{none}
\copyrightyear{2021}
\acmYear{2021}
\acmDOI{10.1145/xxxxxxx.xxxxxxxx}

\copyrightyear{2021}
\acmYear{2021}
\setcopyright{acmlicensed}
\acmConference[CHI '21]{CHI Conference on Human Factors in Computing Systems}{May 8--13, 2021}{Yokohama, Japan}
\acmBooktitle{CHI Conference on Human Factors in Computing Systems (CHI '21), May 8--13, 2021, Yokohama, Japan}
\acmPrice{15.00}
\acmDOI{10.1145/3411764.3445339}
\acmISBN{978-1-4503-8096-6/21/05}

\begin{document}

\title[Mindless Attractor]{Mindless Attractor: A False-Positive Resistant Intervention for Drawing Attention Using Auditory Perturbation}


\author{Riku Arakawa}
\orcid{0000-0001-7868-4754}
\affiliation{%
  \institution{The University of Tokyo}
  \city{Tokyo}
  \country{Japan}
}
\email{arakawa-riku428@g.ecc.u-tokyo.ac.jp}
\authornote{These authors contributed equally and are ordered alphabetically.}
\authornote{Also with ACES Inc., Japan.}

\author{Hiromu Yakura}
\orcid{0000-0002-2558-735X}
\affiliation{
    \institution{University of Tsukuba}
    \city{Tsukuba}
    \country{Japan}
}
\email{hiromu@teambox.co.jp}
\authornotemark[1]
\authornote{Also with Teambox Inc., Japan.}

\renewcommand{\shortauthors}{Arakawa and Yakura}

\newcommand{\tabref}[1]{Table~\ref{tab:#1}}
\newcommand{\figref}[1]{Figure~\ref{fig:#1}}
\newcommand{\secref}[1]{Section~\ref{sec:#1}}
\newcommand{\eqnref}[1]{Equation~\ref{eqn:#1}}

\begin{abstract}
Explicitly alerting users is not always an optimal intervention, especially when they are not motivated to obey.
For example, in video-based learning, learners who are distracted from the video would not follow an alert asking them to pay attention.
Inspired by the concept of \emph{Mindless Computing}, we propose a novel intervention approach, \emph{Mindless Attractor}, that leverages the nature of human speech communication to help learners refocus their attention without relying on their motivation.
Specifically, it perturbs the voice in the video to direct their attention without consuming their conscious awareness.
Our experiments not only confirmed the validity of the proposed approach but also emphasized its advantages in combination with a machine learning-based sensing module.
Namely, it would not frustrate users even though the intervention is activated by false-positive detection of their attentive state.
Our intervention approach can be a reliable way to induce behavioral change in human--AI symbiosis.
\end{abstract}



\maketitle

\section{Introduction}
\label{sec:intro}

For decades, video-based communication has been expected to take over face-to-face communication \cite{DBLP:conf/chi/FishKRR92,DBLP:journals/ijmms/Whittaker95}.
In particular, schools have leveraged video-based learning to provide educational opportunities for distanced students, as massive open online courses have done \cite{DBLP:journals/cacm/Martin12,DBLP:conf/lats/GuoKR14}.
Moreover, the recent COVID-19 pandemic has precipitated the transition to video-based communication for the purpose of preventing infection \cite{Coombs2020,KODAMA2020102172}, especially in the context of education \cite{Kerres2020,Goldschmidt2020}.
However, it has been noted that people often have trouble maintaining their attention in video-based communications \cite{DBLP:conf/chi/KuzminykhR20,DBLP:conf/chi/KuzminykhR20a}, as they can concurrently perform other tasks, like texting or accessing social media using a smartphone \cite{Oeppen2020}. 

Considering the increasing demand for video-based learning, it would be fruitful if computers can help learners pay attention to a video.
Here, recent advances in machine learning techniques have enabled the automatic estimation of a user's attention level from a video of their face \cite{DBLP:conf/icmi/ThomasJ17}.
On the other hand, it is not trivial how to intervene with learners using computers based on the estimation result.
A straightforward approach is to explicitly alert them when they seem to not be paying attention to the video, as Xiao and Wang \cite{DBLP:conf/icmi/XiaoW16} did.
However, unlike the critical situations targeted in conventional studies of alert designs \cite{DBLP:conf/nordichi/CobusMABH18,DBLP:conf/chi/Saint-LotID20}, users of video-based learning systems would not hesitate to ignore such alerts, especially when they are focused on side tasks.
For example, Xiao and Wang \cite{DBLP:conf/icmi/XiaoW16} reported that their intervention approach in their user study was described as unhelpful by some participants who were less motivated.  
In other words, the efficacy of the alerting approach would depend on the user's motivation to actively take part, and such interventions would not be an optimal intervention for inducing behavioral change.

Looking back to the nature of human communications, we often change the tone of our voices intentionally to draw listeners' attention \cite{DBLP:journals/speech/Xu05}.
Based on this observation, we anticipate that we can help learners return their attention to videos by computationally changing the tone of voice during video-based learning situations. 
This approach is inspired by the concept of \textit{Mindless Computing}---behavior-changing technologies that leverage human biases or unconscious behaviors---proposed by Adams et al. \cite{DBLP:conf/huc/AdamsCJC15}.
Given that Mindless Computing does not consume a user's conscious awareness to be effective, Adams et al. \cite{DBLP:conf/huc/AdamsCJC15} stated that it does not rely on the user's motivation, whereas many of the current persuasive technologies have a strong reliance on user motivation and are likely to fail.
In addition, the independence from the user's conscious awareness enables such behavior influencing to work without interfering with the user's main task, which suits our situation (i.e., use during video-based learning).

Furthermore, we argue that this mindless intervention approach has a high affinity with sensing modules based on machine learning techniques.
That is, if we explicitly alert users, they can be distracted and frustrated by misinformed alerts caused by erroneous false-positive detection, which can lead them to ignore the result of a machine learning module \cite{Dietvorst2015,DBLP:conf/chi/De-ArteagaFC20}.
On the other hand, the mindless approach designed based on human nature does not necessarily consume users' conscious awareness, and such negative effects due to false positives can thus be mitigated.

In this paper, we propose a novel intervention approach, \textit{Mindless Attractor}, which computationally leverages the nature of our speech communication, and examine its effectiveness in the case of helping users in video-based learning return their attention to the video.
For this purpose, we first determined its requirements and possible designs so as to reduce the time that users are distracted in a mindless manner.
We then conducted an experiment to confirm that the proposed intervention was effective in helping users refocus their attention without consuming conscious awareness.
We also combined this mindless intervention with a machine learning-based sensing module and evaluated its effectiveness in the context of false-positive detection, in comparison to a conventional alerting approach.
The series of experiments presented the advantages of the proposed approach, especially in combination with machine learning techniques.
Based on the results, we discuss implications for the HCI community, emphasizing the importance of the mindless intervention approach in the era of human--AI symbiosis.

\section{Related Work}
\label{sec:rw}

To situate our work, we first examine previous literature on interaction techniques for video-based learning, particularly those focusing on learners' attention.
We then review conventional alert-based techniques for drawing human attention and discuss why they would not fit our purposes.
We also explore previous studies regarding the nature of human speech communication, as this is a foundation of our mindless approach for drawing users' attention.

\subsection{Attention-Related Interaction Techniques for Video-Based Learning}
\label{sec:rw-interaction}

As mentioned in \secref{intro}, opportunities for video-based communication are increasing, and many interaction techniques have thus been proposed to enhance the experience of such communications.
Some prior studies have proposed interaction techniques centering on the context of participants' attention \cite{DMello2012Gaze,DBLP:conf/lak/SharmaAJD16,DBLP:conf/icmi/XiaoW16}, as it has been pointed out that people often have difficulty maintaining their attention during video-based communication \cite{DBLP:conf/chi/KuzminykhR20,DBLP:conf/chi/KuzminykhR20a}.
These techniques benefit from the significant effort that has been devoted to estimating participants' attentiveness based on visual cues, such as face movement \cite{DBLP:conf/icmi/ThomasJ17}, body postures \cite{DBLP:journals/ejivp/ZaleteljK17}, and gaze \cite{bidwell2011classroom,DBLP:conf/um/HuttMBKBD17,DBLP:conf/ACMse/VeliyathDAHM19}.
They then use the estimation results to enhance learners' performance, for instance in the case of video-based learning, as it is widely acknowledged that learners' attention and engagement are strongly related to their learning performance \cite{Baker2010Better,DMello2012Gaze}.

For example, Gaze Tutor is a gaze-reactive intelligent tutoring system for video-based learning \cite{DMello2012Gaze}.
Using a conventional eye tracker, it estimates the learner's attention level based on gaze direction by applying a simple rule assuming that off-screen gaze patterns imply distraction.
When the system detects that the learner is not focusing on the video, the tutor agent stops the video and alerts them explicitly (e.g., by saying ``Please pay attention'').
Although their experiment showed its effectiveness in reorienting participants' attention, the intervention method left room for improvement, as the authors mentioned in their discussion.
Specifically, they found individual differences in the efficacy of the alert-based intervention, including that some participants never followed the alerts.
Accordingly, the authors noted that alternate intervention approaches, including indirect feedback, could be implemented.
Another example that computationally utilizes the estimated attention level during video-based learning was provided by Sharma et al. \cite{DBLP:conf/lak/SharmaAJD16}.
Similar to Gaze Tutor, their system provided users with direct feedback, such as simple red rectangles on the screen, with the purpose of improving users' attention.

As can be inferred from these studies, previous research has mainly considered explicit alerting as an intervention method for video-based learning.
However, the findings from these studies complement our concern, which is discussed in \secref{intro} based on the results of Xiao et al. \cite{DBLP:conf/icmi/XiaoW16}.
That is, such interventions have a reliance on users' motivation; they may not work effectively when we cannot assume that all users are motivated to change their behavior.
In \secref{rw-drawing}, we will explain why the reliance occurs based on the discussion by Adams et al. \cite{DBLP:conf/huc/AdamsCJC15}, which in turn motivated us to explore a better intervention approach for video-based learning situations.


\subsection{Alerting Techniques for Drawing Human Attention}
\label{sec:rw-drawing}

Drawing users' attention is one of the crucial components of human-computer interaction, not limited to video-based learning.
Many researchers have dealt with a wide range of topics in this area, such as Internet advertisements \cite{DBLP:journals/chb/NettelhorstB12a}, smartphone notifications \cite{Stothart2015The}, and alerting systems \cite{LGuo2002Agricultural}.
Consequently, previous studies have developed many methods suitable for individual situations using diverse perceptual modalities.

One of the most popular strategies is to provide users with visual stimulation.
For example, Red Alert is a visual alerting system which uses a translucent orange-red flash to mask a screen, designed to warn pilots potential collisions in air traffic control \cite{DBLP:conf/chi/Saint-LotID20}.
Audio stimuli have also been favorably employed as a means to alert users.
BBeep is a collision-avoidance system that can emit a beep sound to alert pedestrians around a visually impaired user to clear the way \cite{DBLP:conf/chi/KayukawaHGMSKA19}.
Another strategy is the use of the tactile modality.
BuzzWear is a wrist-worn tactile display to notify users on the go by combining different parameters of the tactile stimulus \cite{DBLP:conf/chi/LeeS10}.
As can be observed in these examples, most systems adopt explicit stimuli to notify users, assuming that they will take action after their attention is drawn to the target.

However, Adams et al. \cite{DBLP:conf/huc/AdamsCJC15} pointed out that such alerting strategies would not be optimal when used within persuasive technologies designed to influence user behavior.
Unlike critical situations (e.g., air traffic control) where it can be expected that users will be motivated to follow an alert from a computer, not all scenarios for inducing behavioral change can assume that users are motivated to do so.
In such cases, an alert that requires the user's conscious awareness and effort to work effectively would likely fail due to lack of motivation or potentially counteract positive aspects of the intervention by frustrating them.
Thus, the authors recommended the Mindless Computing strategy of leveraging human biases or unconscious behaviors, which diminishes reliance on users' conscious awareness.
It also enables a user intervention without interfering with users' ongoing activity, whereas alerting users explicitly can interrupt such activity.
Furthermore, they complimented the advantage of the mindless approach by mentioning that such interventions have long-term effectiveness, even though users are aware of the biases behind the interventions \cite{Wansink2007}.

This point is common to the previous studies for video-based learning in regards to the reliance on learners' motivation, which is mentioned in \secref{rw-interaction}.
That is, as conventional alerting approaches are requiring learners' conscious awareness to be effective, they would have an option not to follow the intervention.
Therefore, for the purpose of helping learners return their attention, we explore a new computational approach that intervenes without consuming their conscious awareness.
This led us to make use of the nature of human speech communication.


\subsection{Speech Communication Techniques for Drawing Human Attention}
\label{sec:rw-human}

Speech is one of the most natural modalities of human communication.
It consists not only of linguistic aspects but also of paralinguistic aspects, such as pitch, volume, and speed, which play an important role in conveying nuance or emotion \cite{Trager1958ParalanguageA}.
Though the use of paralinguistic aspects is a natural habit that does not necessarily require our conscious processes \cite{9027235279}, it is also a common practice to intentionally create changes in such paralinguistic parameters while speaking so as to draw listeners' attention \cite{Toastmasters}.
The relationship between speech parameters and their effects in terms of drawing attention has generated considerable research interest in understanding human speech communication.
For example, Xu \cite{DBLP:journals/speech/Xu05} confirmed that an increase in pitch when starting a new topic can draw listeners' attention.
Moreover, a similar effect of drawing attention has also been observed in infants hearing the speech of their mothers, who naturally vary their pitch \cite{Sullivan1983TheEffects}.
The idea that humans unconsciously respond to paralinguistic cues is further supported by Zatoree and Gandour \cite{Zatorre2007Neural}, who verified that human neural mechanisms are sensitive to such spectral and temporal acoustical properties.

Based on these results, we speculate that leveraging this nature of human speech communication by computationally varying speech parameters can draw listeners' attention in a natural manner.
More specifically, if a person losing their attention to a video hears speech with altered pitch or volume, they will naturally respond to such a change, regardless of their motivation to pay attention.
Such an intervention approach is in line with the concept of Mindless Computing \cite{DBLP:conf/huc/AdamsCJC15} and thus is expected to work without depending on users' motivation.
In the following section, we further elaborate on the rationale for our design of using alterations of human speech to draw attention in video-based learning situations.

\section{Mindless Attractor}
\label{sec:mindless}

In this paper, we propose \textit{Mindless Attractor} for the purpose of helping users in video-based learning situations return their attention to the video.
Inspired by the concept of Mindless Computing \cite{DBLP:conf/huc/AdamsCJC15}, it leverages the nature of speech communication to intervene with users.
In this section, we present the details of Mindless Attractor, starting by discussing why the mindless approach should be considered and what requirements should be fulfilled.

\subsection{Why Mindless?}
\label{sec:mindless-why}

As we stated in \secref{intro}, our research aim is to support video-based learning, given the growing demand for it, by establishing a suitable computational intervention for users who are not paying attention to the video.
The difficulty is that we cannot assume all users to be highly motivated to follow such an intervention for maintaining attention, which we mentioned in \secref{rw-drawing} as the reason that conventional alerting approaches would not be suitable.
Thus, we need to consider an intervention approach that does not rely on users' motivations.
In addition, even when a user is not focusing on the video, intervention approaches that interrupt the user should be avoided since such approaches might lead them to miss subsequent content.

These points led us to adopt an approach based on Mindless Computing \cite{DBLP:conf/huc/AdamsCJC15} that leverages human biases or unconscious behaviors to induce behavioral change.
Since such an intervention approach does not consume the user's conscious awareness to be effective, it is considered less reliant on their motivation to pay attention.
Moreover, it enables us to design a less interruptive intervention than explicit alerts, as Adams et al. \cite{DBLP:conf/huc/AdamsCJC15} confirmed that their mindless approach using auditory feedback could influence people's behavior when talking without annoying them.

Furthermore, we presume that the mindless approach will reveal a new advantage when integrated with a sensing module based on machine learning techniques, as mentioned in \secref{intro}.
More specifically, although machine learning systems enable various sensing scenarios, humans tend to evaluate such systems' mistakes more severely than human mistakes \cite{Dietvorst2015}. 
In addition, the trust that machine learning systems lose as a result of their failure is usually greater than the trust they gain from their success \cite{DBLP:conf/iui/YuBTCZC17}.
Consequently, people often become less hesitant to override outputs from machine learning systems after seeing their failures \cite{DBLP:conf/chi/De-ArteagaFC20}.
Moreover, it has been suggested that people with a high level of cognitive load will have less trust in interactions with machine learning systems \cite{DBLP:conf/interact/ZhouALC17}.
These discussions imply the risk posed by the false-positive detection of the sensing module in intervening with users---that is, mistakenly alerting them in an explicit manner during video-based learning situations would frustrate them and lead them to disregard the alerts.
On the other hand, since the mindless approach does not consume conscious awareness, unlike the alerting approach, it might mitigate the negative effects caused by false positives.

We therefore suppose that the mindless approach would be suitable as an intervention in the context of video-based learning.
In particular, we believe that this is a plausible solution to the current situation where effective interventions for video-based learning have not been well investigated, as discussed in \secref{rw-interaction}.

\subsection{Designing Mindless Attractor}
\label{sec:mindless-design}

To design the mindless approach leveraging human biases or unconscious behaviors, we exploited the nature of human speech communication.
Our design is based on the following requirements we considered in view of using the mindless approach in video-based learning situations.
\begin{description}
    \item[Avoid interruption due to interventions.] Considering that video-based learning is sometimes delivered in the form of live streams or in a synchronous manner \cite{BATURAY2015427}, interrupting users due to interventions should be avoided, as it can cause them to miss information and counteract our aim of helping them pay attention. This requirement is one reason to eliminate the use of alerting approaches, as we discussed their interruptive aspect in \secref{mindless-why}.
    \item[Use a modality that users will not neglect.] To intervene with users who are not paying attention to the video, it is important to use a modality that is always reachable for users. In this regard, though it is possible to leverage human perceptual bias to design the mindless approach by showing something on a display, this would not be suitable because the user can take their eyes off the display, especially when performing other tasks using a smartphone \cite{Oeppen2020}. On the other hand, it seems more unlikely that the user would not hear the audio due to muting it while in video-based learning situations.
    \item[Function without external devices.] Though the use of external devices would extend the range of possible interventions, such as using a tactile stimulus \cite{DBLP:conf/chi/LeeS10}, it raises an additional cost to utilize the interventions. Therefore, it is desirable to design an intervention that could be integrated into video-based learning situations without requiring external devices.
\end{description}

As we reviewed in \secref{rw-human}, it has been suggested that humans unconsciously respond to paralinguistic cues in speech, such as a change in pitch, volume, and speed.
In our case, we considered perturbing the pitch or volume of the voice in the video to help users refocus their attention.
We did not use speed because it would be difficult to maintain time-series consistency when video-based learning is conducted in a synchronous manner (e.g., live lectures \cite{BATURAY2015427}).

In addition, the perturbation is enabled and disabled repeatedly when the user is seemingly not paying attention to the video, as Adams et al. \cite{DBLP:conf/huc/AdamsCJC15} emphasized the importance of cues to trigger different perceptions and sensations in designing mindless approaches.
Otherwise, if we activated the perturbation once when the user became distracted and kept it thereafter, the user would have less opportunity to refocus their attention as they became acclimated to the changed pitch or volume.

\subsection{Implementation}
\label{sec:mindless-implementation}

We used Python and PyAudio\footnote{\url{https://people.csail.mit.edu/hubert/pyaudio/docs/}} to perturb the audio signal in real time.
The audio signal was captured in 16~kHz and the perturbation process was activated each 1/16 sec to ensure that the perturbed signal was delivered without significant delay.
The pitch shift was performed using a library named rubberband\footnote{\url{https://github.com/breakfastquay/rubberband}} through time-shifting and resampling the signal via Fourier transform.
The volume change was performed by directly multiplying the waveform double or halve.
Our source code is publicized at a GitHub repository\footnote{\url{https://github.com/hiromu/MindlessAttractor}}.

In addition, as we mentioned in \secref{intro} and \secref{mindless-why}, our mindless intervention approach is expected to incorporate a sensing module that monitors users' behavior and detects when they are distracted.
The detailed implementation of the sensing module is later explained in \secref{exp2-material}.

\section{Hypotheses}
\label{sec:hypotheses}

Up to this point, we have introduced Mindless Attractor, which is designed as an intervention for users during video-based learning that incorporates a sensing module based on machine learning techniques.
It computationally perturbs the pitch and volume of the voice in the video in real time to refocus users' attention when they seem to be distracted from the video.
Our design rationale for the proposed approach, which we discussed in \secref{mindless-design}, imposes the following hypotheses, which need to be verified to ensure the validity and effectiveness of the proposed approach.

First, as we discussed in \secref{mindless-why}, our proposal is based on the concept of Mindless Computing \cite{DBLP:conf/huc/AdamsCJC15} so as to ensure that the intervention works without relying on user motivation and without interrupting users.
To satisfy these points, we should examine whether Mindless Attractor can influence users' behavior in a mindless manner, i.e., without consuming their conscious awareness.
\begin{quote}
    H1: Mindless Attractor is an effective means to refocus the attention of users in video-based learning situations without consuming their conscious awareness.
\end{quote}

If H1 holds, we have two choices for inducing behavioral change in users (i.e., drawing their attention back to the video): alerting users in an explicit manner or intervening in a mindless manner.
Here, as we discussed in \secref{mindless-why}, we expect that the proposed approach will be favored over alerting approaches when combined with a machine learning-based sensing module that detects when users are losing attention.
More specifically, the fact that such a sensing module may produce false positives implies the risk of mistakenly intervening in users, which can be annoying when we alert them explicitly.
Thus, we posit our second hypothesis:
\begin{quote}
    H2: Mindless Attractor is not only an effective means to refocus users' attention but is also preferred by users when combined with a machine learning-based sensing module, while the alerting approach is not accepted.
\end{quote}

If these hypotheses are supported, we can pave the way for intervening with users in real time to support their participation during video-based learning.
With this motivation, we evaluated these hypotheses by conducting a series of experiments.

\section{Experiment I: Evaluation of H1}
\label{sec:exp1}

\subsection{Design}

To evaluate H1, we conducted an experiment that replicated video-based learning situations.
We used a within-participant design comparing a treatment condition using Mindless Attractor with a control condition that did not intervene in participants.
Then, H1 is supported if the following two points are confirmed: Mindless Attractor helps participants refocus their attention, and Mindless Attractor does not consume participants' conscious awareness.

\subsection{Measure}

We prepared two measures corresponding to the above two points to be confirmed: recovery time and cognitive workload.

\subsubsection{Recovery Time}

This metric indicates the time that it took for participants to return their attention to the video after losing focus.
If Mindless Attractor helps participants refocus their attention, the time that they are distracted should be shortened in comparison to the case in which no intervention was taken.

To compute this metric, we collected human annotations for each participant denoting whether the participant was paying attention or not.
As we explain in the detailed procedure description in \secref{exp1-procedure}, an experimenter observing the state of the participants annotated in real time so that the recovery time could be calculated later.

\subsubsection{Cognitive Workload}

This metric was used to evaluate whether Mindless Attractor consumed the participants' conscious awareness or not.
Measuring cognitive workload is common in the previous studies proposing alerting approaches \cite{DBLP:conf/chi/Saint-LotID20,DBLP:conf/chi/LeeS10}.
Whereas they aimed to show that their proposed approaches exhibited lower workload compared to other possible approaches, we compared the metric between the control and treatment conditions.
If the cognitive workload in the treatment condition is not significantly different from that in the control condition, it suggests that Mindless Attractor does not consume participants' conscious awareness.
In our study, we used the NASA-TLX questionnaire \cite{Hart1988,Byers1989} to measure cognitive workload, in the same manner as the previous studies \cite{DBLP:conf/chi/Saint-LotID20,DBLP:conf/chi/LeeS10}.

We note that it would be possible to evaluate whether Mindless Attractor consumes the participants' conscious awareness by asking them whether they noticed the perturbation.
However, to do so, we would need to conceal from the participants that they would be subject to an intervention, which would create an unrealistic situation if we consider the practical applications of the proposed approach.
More specifically, it is unlikely that users in video-based learning situations would be subject to interventions without opt-in consent; that is, they would use Mindless Attractor of their own accord to focus on videos or at least would be notified about the possibility of the intervention.
In addition, as we mentioned in \secref{rw-drawing}, Adams et al. \cite{DBLP:conf/huc/AdamsCJC15} explained that the mindless approaches work regardless of whether a user knows their mechanisms or not, as they do not depend on the user's conscious awareness.
Thus, we used this measure based on NASA-TLX and also notified the participants beforehand that they would be subject to interventions.

\subsection{Material}
\label{sec:exp1-material}

To replicate a video-based learning situation, we prepared a video recording of a 30-minute lecture on urban sociology.
As this experiment was conducted remotely, the video was presented to the participants using the screen-sharing function of Zoom\footnote{\url{https://zoom.us/}}.

By following the implementation we described in \secref{mindless-implementation}, we also prepared a client software that modifies Zoom's audio output to perform our intervention.
This software captures and perturbs the audio output in real time when it receives an activation command from a control server via WebSocket.
Here, we conducted a pilot study in the same manner as Adams et al. \cite{DBLP:conf/huc/AdamsCJC15} to find the best parameters for intervening without causing distractions.
Consequently, we implemented four perturbation patterns: halving or doubling the volume and lowering or raising the pitch by one tone.
The software then activates one of the four patterns randomly so as to enable the comparison of their effectiveness for helping the participants refocus their attention.
Since Zoom automatically removes noises and extracts voices, we confirmed that our na\"{i}ve implementation of pitch shifting based on fast Fourier transform would be sufficient for the purposes of this experiment.

We further prepared an experimenter console in the control server to record annotations concerning whether the participant was paying attention or not.
The console was implemented to enable sending the activation and deactivation command to the client software when the participant started to divert their attention from the video and refocused their attention, respectively.

\subsection{Participants}

This experiment involved 10 participants, three of whom were female.
They were recruited via online communication in a local community where over 100 university students gather.
As described later in \secref{exp1-procedure}, our experimental procedure required participants to be observed by a remote experimenter so that their state of attention could be annotated.
Therefore, we asked them to prepare a PC with a webcam in a quiet room as well as to enable their faces to be captured.

\subsection{Procedure}
\label{sec:exp1-procedure}

Each participant underwent one session of watching the 30-minute video using a computer connected over Zoom, as we mentioned in \secref{exp1-material}.
To replicate the usual situation of video-based learning, in which learners have some reasons to watch the video, we told participants in advance that they would be asked to write a few sentences summarizing the video.
At the same time, we asked them to bring their smartphones and told them that we would not prohibit the use of smartphones so that they could be distracted as usual \cite{Oeppen2020}.

\begin{figure*}
    \centering
    \includegraphics[width=0.8\linewidth]{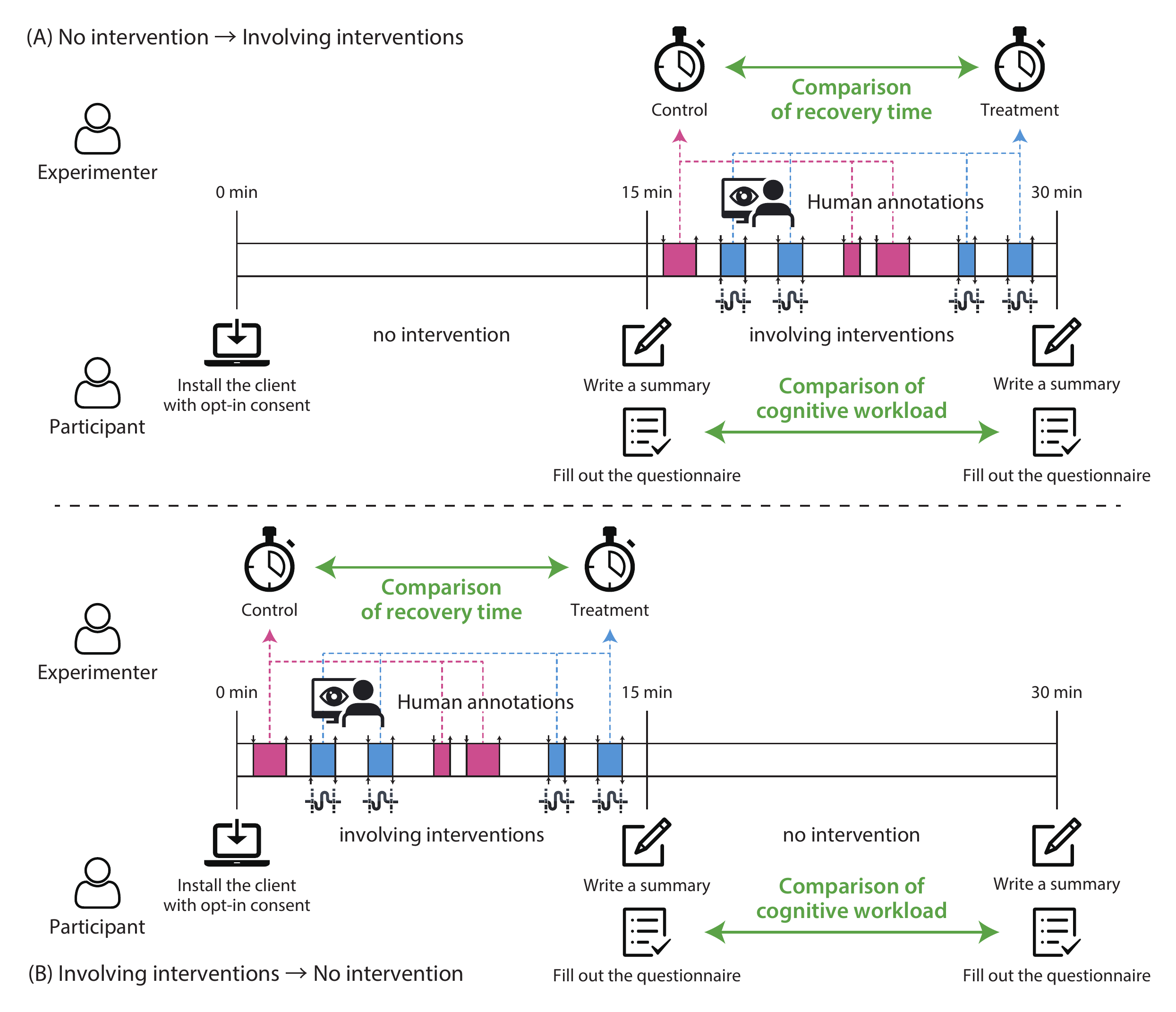}
    \caption{Example illustration of the procedure for our first experiment. (A) Half of participants first experienced the part with no intervention and then experienced the part involving interventions, and (B) the others followed the reversed order.}
    \Description{Example illustration of the procedure for our first experiment. After each participant installed the client with opt-in consent at the beginning, they experienced two parts: one with no intervention and another involving interventions. In the part involving interventions, an experimenter annotated whether the participant was paying attention to the video. Half of participants followed this order, and the others first experienced the part involving interventions and then experienced the part with no intervention.}
    \label{fig:exp1-procedure}
\end{figure*}

As depicted in \figref{exp1-procedure}, each session was divided into two parts of 15 minutes each: one with no intervention and another involving interventions.
To normalize the order effect, we balanced the order of the two parts: five participants first experienced the part with no intervention, and the others first experienced the part involving interventions.
After each part, the participant was asked to write a summary and fill out the questionnaire measuring cognitive workload.
Note that these two parts do not correspond to the control and treatment conditions, as explained in the following paragraphs.

In the part involving interventions, an experimenter observed the state of a participant, including their use of smartphones, and annotated whether they were paying attention to the video or not.
When the experimenter pressed a button on the experimenter console to record the timestamp at which the participant diverted their attention from the video, the console assigned either the control or treatment condition with a 50\% probability of each.
Note that the selected condition was concealed from the experimenter in order to avoid the experimenter bias in the annotations.
If the treatment condition was assigned, the console sent the activation command to the client, and the client then repeatedly enabled and disabled one of the four perturbation patterns every 3 seconds, as explained in \secref{mindless-design}.
This intervention continued until the client received the deactivation command indicating that the experimenter pressed another button to record the participant's recovery from the distraction.
On the other hand, if the control condition was assigned, no command was sent to the client.
Consequently, based on the assigned conditions and the recorded timestamps, the recovery time could be calculated and compared.

The other part (with no intervention) was prepared to evaluate the cognitive workload.
We compared its cognitive workload score with that of the part involving interventions, which were activated on a random basis.
If the intervention did not consume the participant's conscious awareness, the scores of the two parts would not be significantly different.

In addition, at the end of the session, we asked the participants for their comments about their feelings or anything they noticed.
In total, the entire session took about an hour to complete.

\subsection{Results}
\label{sec:exp1-results}

\begin{table*}
    \caption{Comparison of the recovery time and cognitive workload score between the control and treatment conditions. The treatment condition involved the mindless intervention.}
    \label{tab:exp1}
    \begin{minipage}{\columnwidth}
        \centering
        \begin{tabular}{lrrr}
            \toprule
            \multicolumn{1}{c}{Measure} & \multicolumn{1}{c}{Treatment} & \multicolumn{1}{c}{Control} & \multicolumn{1}{c}{$p$-value} \\
            \midrule
            Recovery time               & $17.71~s$ ($\pm 10.52~s$)     & $32.25~s$ ($\pm 16.92~s$)   & < 0.0001                      \\
            Cognitive workload          & $26.00$ ($\pm 10.32$)         & $27.00$ ($\pm 9.13$)        & 0.5212                        \\
            \bottomrule
        \end{tabular}
    \end{minipage}
\end{table*}

\subsubsection{Recovery Time}

\begin{figure*}
    \centering
    \includegraphics[width=0.92\linewidth]{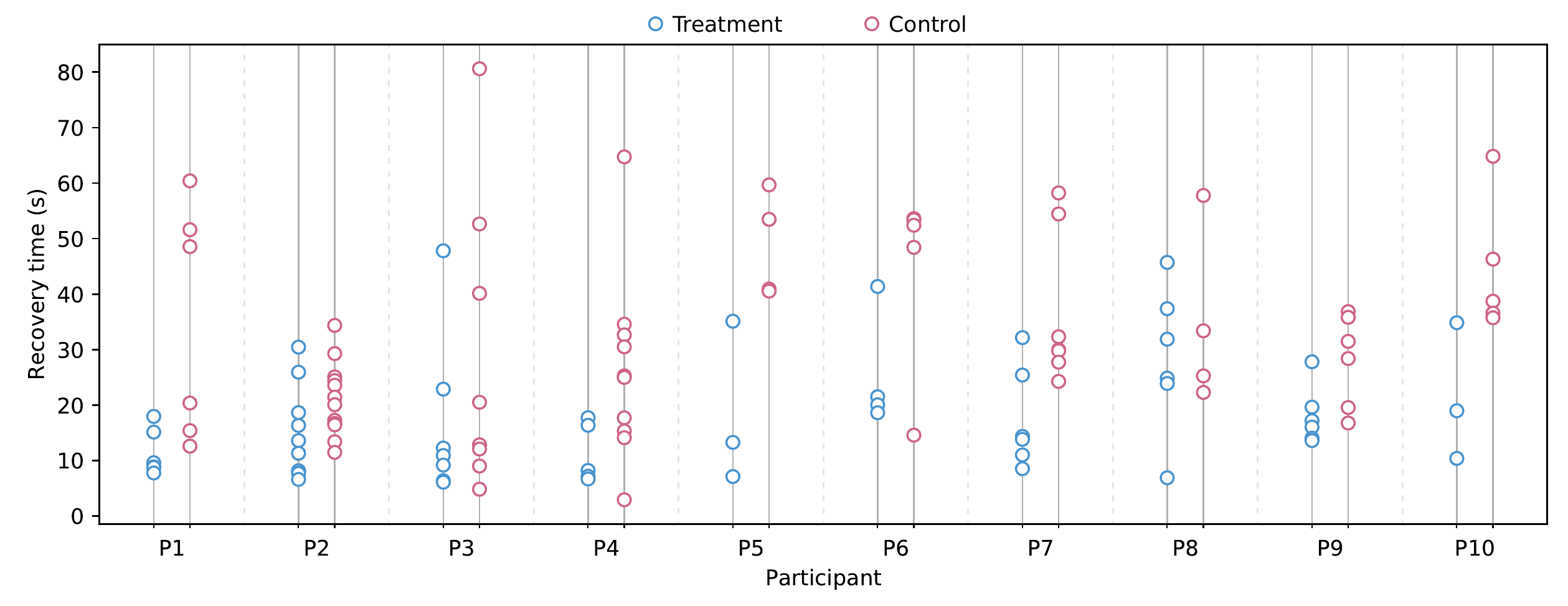}
    \caption{Distribution of the recovery time across each participant and the experimental conditions.}
    \Description{Distribution of the recovery time across each participant and the experimental conditions. The participants tended to spend more time recovering their attention in the control condition.}
    \label{fig:exp1-result}
\end{figure*}

As shown in \tabref{exp1}, the proposed intervention significantly shortened the recovery time according to the unpaired $t$-test (Cohen's $d = 1.0044$, $p < 0.0001$).
The distribution of the recovery time is shown in \figref{exp1-result}, which also confirms this reduction.
This result supports that Mindless Attractor helped participants refocus their attention.

We also investigated which of the four perturbation patterns (i.e., halving or doubling the volume and lowering or raising the pitch by one tone) effectively helped participants refocus their attention.
We examined the last perturbation pattern before each time the participant returned their attention and counted their occurrence, as shown in \tabref{exp1-patterns}.
This examination is based on our assumption that the intervention just before the participant's attention returned is the cause of the change in the participant's state.
According to the $\chi^2$-test comparing with the total occurrence, the results were not significantly different in that each pattern equally helped participants recover their attention (Cramer's $V = 0.1220$, $p = 0.2794$).
In other words, we can conclude that there was no significant difference in the effectiveness of the four perturbation patterns.

\begin{table*}
    \caption{Occurrence of the four perturbation patterns that were executed just before participants returned their attention. The comparison with the total occurrence suggests that there was no significant difference in effectiveness ($p = 0.2794$).}
    \label{tab:exp1-patterns}
    \begin{minipage}{\linewidth}
        \centering
        \begin{tabular}{lcccc}
            \toprule
            Perturbation                          & Halve the volume    & Double the volume  & Lower the pitch     & Raise the pitch     \\
            \midrule
            Occurrence just before                & \multirow{2}{*}{19} & \multirow{2}{*}{7} & \multirow{2}{*}{14} & \multirow{2}{*}{16} \\
            participants returned their attention &                     &                    &                     &                     \\[0.4em]
            Total occurrence                      & 50                  & 47                 & 50                  & 55                  \\
            \bottomrule
        \end{tabular}
    \end{minipage}
\end{table*}

\subsubsection{Cognitive Workload}

We also could not find a significant difference in participants' cognitive load scores according to the paired $t$-test (Cohen's $d = 0.2110$, $p = 0.5212$), as presented in \tabref{exp1}.
That is, it is suggested that Mindless Attractor did not consume participants' conscious awareness or at least did not negatively affect participants' cognitive load by consuming their conscious awareness.
Thus, in combination with the effect on the recovery time, H1 was supported.

\subsubsection{Comments}

We also examined the comments that the participants wrote at the end of the experiment.
At first, we realized that three participants mentioned that they did not notice any intervention, although they were informed of the intervention beforehand.
Interestingly, the recovery time for these three participants also showed a significant difference (Cohen's $d = 0.8105$, $p = 0.0122$) between the treatment ($15.88~s$ on average) and control ($28.34~s$ on average) conditions.
Thus, it is suggested that the mindless approach worked even when it was not noticed by participants, further supporting that Mindless Attractor did not consume the participants' conscious awareness.
This point not only corroborates H1 but also shows consistency with the discussion by Adams et al. \cite{DBLP:conf/huc/AdamsCJC15}.

It was also interesting that, although five participants mentioned that they noticed the changes in volume, no participant recognized the changes in pitch.
That is, although no significant difference was found between the effectiveness of the four perturbation patterns in \tabref{exp1-patterns}, their noticeability varied, suggesting further room for investigation.

Nevertheless, no participants regarded the mindless intervention as disruptive or annoying; rather, two participants made positive comments about it:
\begin{quote}
    I found it useful because it naturally brought my attention back to the video when I thought something might have changed in the speech. (P1)
\end{quote}
\begin{quote}
    It was nice as it made me feel like...the computer was recommending me to concentrate, rather than warning me. (P4)
\end{quote}
In particular, the latter comment suggested that the mindless approach can mitigate the negative effect that might be caused by false-positive detection when combined with a machine learning-based sensing module.
These results motivated us to conduct a second experiment to evaluate this possibility, as discussed in \secref{hypotheses} when posing H2.

\section{Experiment II: Evaluation of H2}
\label{sec:exp2}

\subsection{Design}

To evaluate H2, we conducted an experiment that replicated a video-based learning situation in the same manner as \secref{exp1}.
However, in this case, we combined a machine learning-based sensing module rather than manually activating interventions and compared the effects of the mindless approach and the alerting approach.
Here, we used a within-participant design over three conditions: mindless, alerting, and control (no intervention).
We added the control condition to confirm that the proposed approach was at least effective in contributing to refocusing users' attention as an automated system controlled by a machine learning-based sensing module.
H2 is thus supported if the following two points are confirmed: Mindless Attractor helps participants refocus their attention, and participants favor Mindless Attractor over the alerting approach.

\subsection{Measure}

Similar to the first experiment, we measured time with regards to whether participants were paying attention.
However, we introduced a different approach for evaluating the time factor, i.e., total distracted time instead of the recovery time.
In addition to this, we introduced a measure for behavioral intention.

\subsubsection{Total Distracted Time}

Although we have confirmed that Mindless Attractor can help participants return their attention, it is desirable to investigate whether the total time that they are distracted during video-based learning is decreased.
In other words, it may be possible that, though the mindless approach shortened the recovery time, the participants were distracted more frequently, especially when the mindless approach was combined with a machine learning-based sensing module having a risk of false positives.

To compute this metric, we collected human annotations for each participant, as we did in \secref{exp1}, and aggregated the duration when the participants were not paying attention.
If the total distracted time in the mindless condition is significantly shorter than in the control condition, it is suggested that Mindless Attractor can make users more likely to pay attention, even in combination with a machine learning-based sensing module.

It should be noted that, due to the false negatives of such a sensing module, there would be a case when the intervention is not triggered even when the participant is actually losing their attention and a case when the intervention is deactivated before the participant refocus.
Therefore, calculating the recovery time as in \secref{exp1-results} is not appropriate in this second experiment, further rationalizing the introduction of the total distracted time as a different metric.

\subsubsection{Behavioral Intention}

This metric was prepared to evaluate whether the mindless approach was favored over the alerting approach.
The concept of behavioral intention is guided by the Technology Acceptance Model \cite{Davis1989Perceived}, which explains users' attitudes towards technologies, and is frequently used to evaluate how likely individuals are to use the technologies.
We used the questionnaire to measure behavioral intention in the same manner as the previous studies \cite{Venkatesh2003User}.
If this score in the mindless condition is significantly better than that in the alerting condition, we can confirm that Mindless Attractor can be favored over the alerting approach, especially when it works as an automated system with a sensing module.

\subsection{Material}
\label{sec:exp2-material}

Similar to our first experiment, we prepared a video recording of a 30-minutes lecture on social sciences.
The experiment was conducted remotely and the video was presented using Zoom's screen-sharing function, as in the first experiment.
However, in this second experiment, we developed a system that automatically detected the status of participants' attention.
To implement this sensing module, we followed previous studies that estimated participants' attentiveness based on their visual cues, which we reviewed in \secref{rw-interaction}.
Specifically, we analyzed the video stream of face images of each participant by leveraging machine learning techniques that can detect their head pose in real time.
If the module detected that the participant was looking off the screen, the system judged that the participant was failing to pay attention to the video lecture, and activated an intervention.

\begin{figure}
    \centering
    \includegraphics[width=0.99\linewidth]{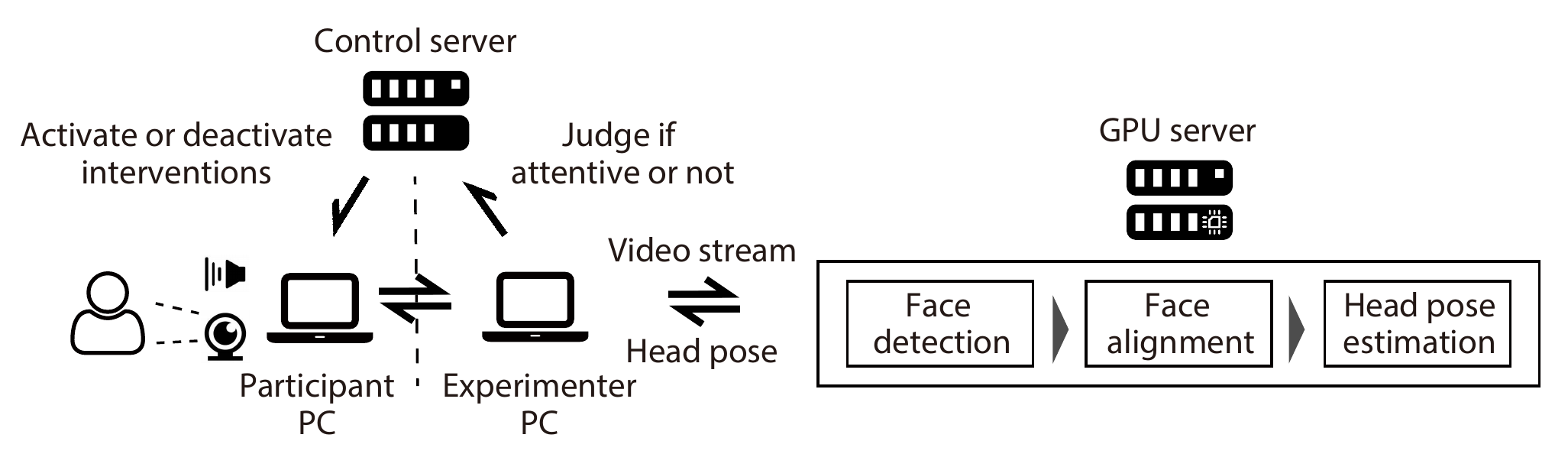}
    \caption{Architecture of the entire system we implemented for the second experiment.}
    \Description{Architecture of the entire system we implemented for the second experiment. From each frame of the video capturing the participant’s face, a GPU server detects the face position, calculates the face alignment, and estimates the head pose. Based on the head pose, an experimenter PC activates interventions via a control server.}
    \label{fig:sensing}
\end{figure}

\figref{sensing} illustrates how the system processed the video streams of participants and intervened in them.
Videos were processed in a frame-by-frame manner.
First, a human face was detected and located in each frame using a deep learning model, RetinaFace \cite{DBLP:conf/cvpr/DengGVKZ20}.
We used this model because it achieves state-of-the-art performance and its pretrained model is publicly released.
Face alignment was then performed to obtain facial keypoints using a deep learning model proposed by Fang et al. \cite{DBLP:conf/iccv/FangXTL17} that is also known to estimate keypoints with high accuracy.
Finally, based on the estimated facial keypoints, the head pose was calculated by solving a perspective-n-point problem.
These calculations were performed using a dedicated computation server with an NVIDIA V100 Tensor Core GPU.

Next, the estimated head pose was passed to the experimenter's PC, a conventional laptop with a 2.2~GHz Intel Core i7 processor.
This PC checked whether the passed head direction was off-screen or not.
The experimenter had conducted a calibration process beforehand to calculate the threshold for this judgment, in which participants were asked to track a red circle that appeared and moved along the edge of the screen.
Participants were told to track the circle by moving their head, i.e., not following it only by moving their gaze.
We then calculated the maximum head rotations for each direction (top--down and left--right) and regarded them as the range where the head is toward the screen.
In other words, when the estimated head pose was out of this range, then the system judged that the participant was looking off the screen, and thus, losing their attention.

While the participants were watching the video, changes in their state--i.e., whether they were looking at the screen or not--were shared with another control server maintaining a WebSocket connection with the client software.
The control server then correspondingly sent activation or deactivation commands in the same manner as the first experiment.
All of the above processes were performed in real time with a frame rate of 15 FPS.

In addition to the sensing module, we implemented an intervention to explicitly alert users in the client software, to be compared with our proposed approach.
In this case, the client software played a short beep for 0.1 seconds, which followed the previous study's use of a beep alert \cite{DBLP:conf/chi/KayukawaHGMSKA19}, rather than perturbing the audio output.
Once the alert was activated, it replayed the same beep every 3 seconds until it received the deactivation command, in the same manner as the mindless condition.

\subsection{Participants}

This experiment involved 20 participants, five of whom were female.
They were recruited in the same manner as we did in the first experiment.
Eight of the participants participated in our first experiment, which had been held at least two weeks before this experiment.
The participants were asked to prepare a PC in a quiet room and to enable their faces to be captured with a webcam, as in the first experiment.

\subsection{Procedure}
\label{sec:exp2-procedure}

Similar to the first experiment, each participant experienced a session of watching the 30-minute video using a computer connected over Zoom.
As before, we told participants in advance that they would be asked to write a few sentences summarizing the video and also allowed them to bring and use their smartphones in the session.

\begin{figure*}
    \centering
    \includegraphics[width=0.8\linewidth]{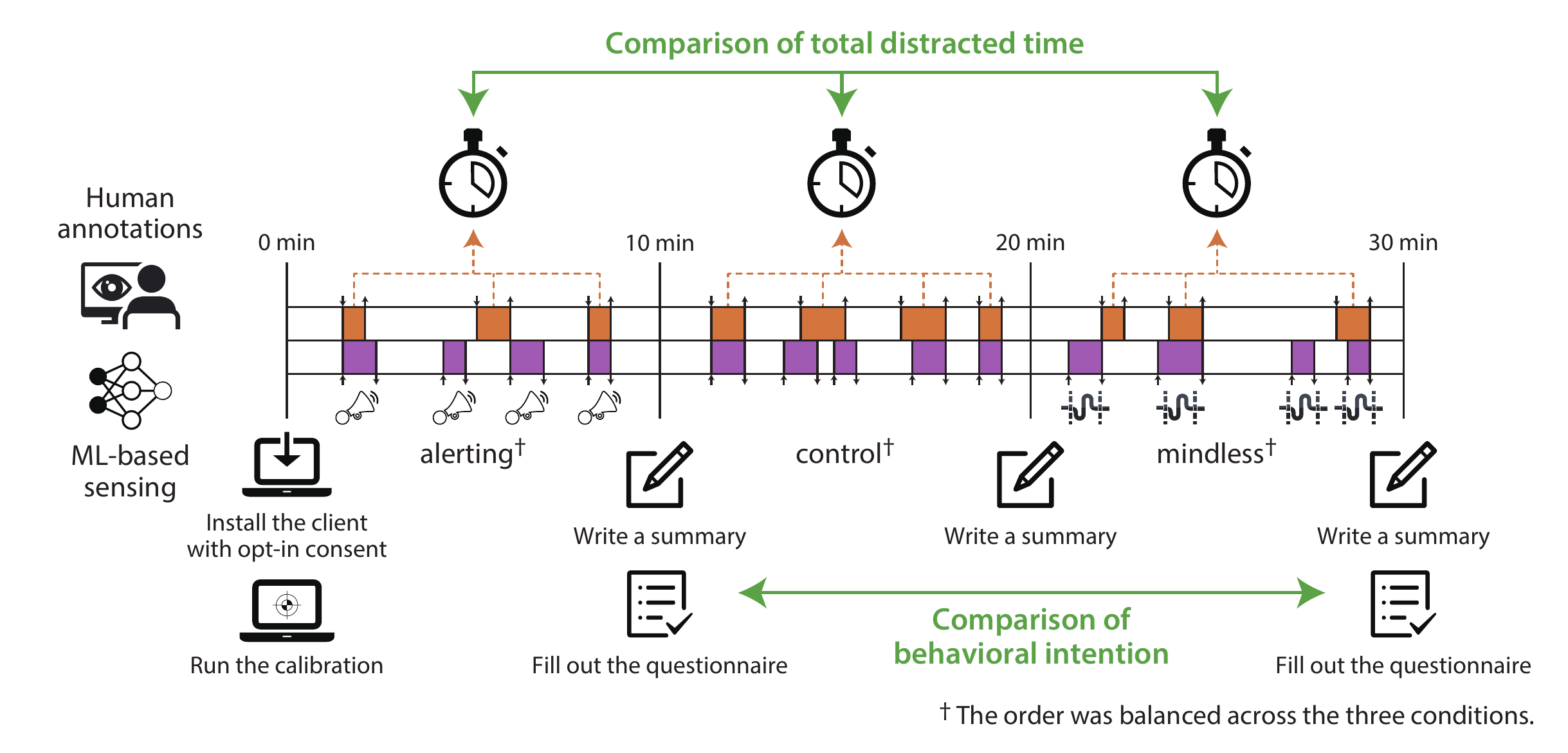}
    \caption{Example illustration of the procedure for our second experiment. Each participant was randomly assigned to one of six possible orders of the three conditions.}
    \Description{Example illustration of the procedure for our second experiment. After each participant installed the client and performed the calibration process at the beginning, they experienced three conditions: mindless, alerting, and control. Each participant was randomly assigned to one of six possible orders of the three conditions.}
    \label{fig:exp2-procedure}
\end{figure*}

As illustrated in \figref{exp2-procedure}, each session consisted of three parts lasting 10 minutes each: one with no intervention, another with the mindless approach, and a third with the alerting approach.
The order of these three parts was automatically randomized among participants, as we will describe later in this section.
After each session, participants were asked to write the summary.
They were also asked to fill out the questionnaire measuring behavioral intention when they finished a part with either the mindless or alerting approach.
We compared the scores between the two conditions to examine which approach participants favored.

Before starting the first session, the experimenter performed a calibration process to determine the threshold for whether the participant's head pose was out of the screen, as described in \secref{exp2-material}.
The experimenter explained that the participants should not move their PC until the entire process was complete and advised them to find a comfortable position before the calibration process started.

In each of the three parts, the experimenter manually annotated whether the participant was paying attention to the video lecture, similar to the first experiment.
To avoid bias, the experimenter was blind to which of the three conditions had been applied to the participant.
Specifically, the control server (see \figref{sensing}) decided the order of conditions in each session, and the experimenter did not have access to this information until the session ended.
The obtained annotations were used to calculate the total distracted time for each part.

In addition, our developed machine learning-based sensing module triggered interventions to the participants in either the alerting or mindless condition, as described in \secref{exp2-material}.
In the alerting condition, participants were exposed to the beep sound when the system judged that they were losing attention, whereas they were exposed to perturbations in the speech in the mindless condition.
In the control condition (i.e., that with no intervention), the client system did not intervene.
In each part, the sequence of the system's judgment was recorded along with timestamps, which we later used to assess the accuracy of the sensing module by comparing it with the human annotations.

Finally, at the end of the session, we asked the participants for their comments about their feelings or anything they noticed.
In total, the entire session took about an hour to complete.

\subsection{Results}
\label{sec:exp2-results}

\subsubsection{Sensing Accuracy}

\begin{table}
    \caption{Confusion matrix between the human annotations and the detection results of the machine learning-based module in regard to participants' attentive state.}
    \label{tab:exp2-accuracy}
    \begin{minipage}{\columnwidth}
        \centering
        \begin{tabular}{c@{\hspace{1\tabcolsep}}l@{\hspace{1\tabcolsep}}r@{\hspace{1\tabcolsep}}r}
            \toprule
                                               &            & \multicolumn{2}{c}{Detection result}                           \\
                                               &            & \multicolumn{1}{c}{Attentive} & \multicolumn{1}{c}{Distracted} \\
            \midrule
            \multirow{2}{*}{
            \shortstack{Human \\ annotations}
            }                                  & Attentive  & 435.4~min (68.5~\%)           & 78.0~min (12.3~\%)             \\
                                               & Distracted & 51.4~min (8.1~\%)             & 70.7~min (11.1~\%)             \\
            \bottomrule
        \end{tabular}
    \end{minipage}
\end{table}

We first examined the accuracy of our machine learning-based sensing module in detecting participants' attentive state.
We compared the human annotations and the detection results of the module and obtained \tabref{exp2-accuracy}.
Though our aim is not to develop a detection system, the accuracy across all the participants was 79.6~\%, which was relatively close to the previous study \cite{DBLP:conf/icmi/ThomasJ17} that achieved the accuracy of 82--85~\% using only head pose.
We note that the accuracy varied among participants (64.9--93.0~\%), which implies that some environmental factors (e.g., the distance to camera or lighting conditions) might largely affect the detection results.
At the same time, the sensing module exhibited a lot of false-positive detection, as its precision was 47.6~\%, which suited our aim to investigate the effect of Mindless Attracter while having a risk of false positives.

\subsubsection{Total Distracted Time}

\begin{figure*}
    \begin{minipage}[t]{0.33\linewidth}
        \centering
        \includegraphics[width=0.91\textwidth]{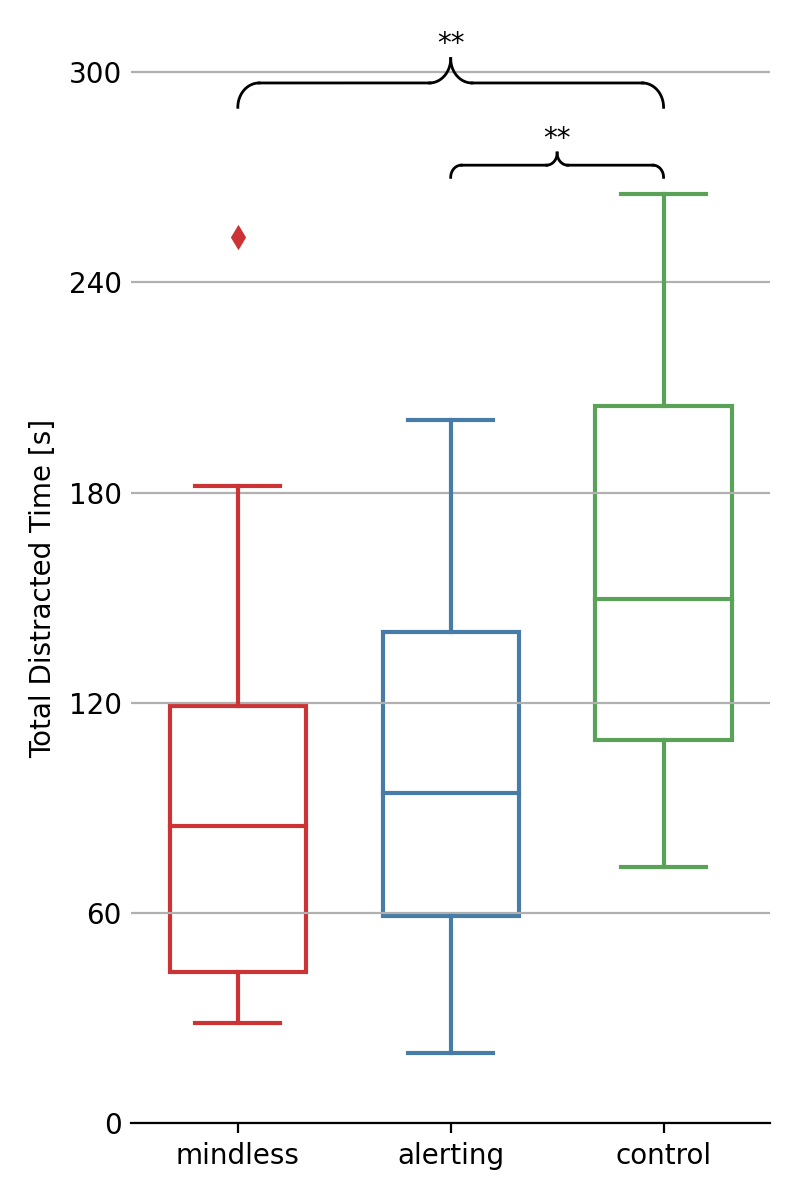}
        \caption{Comparison of participants' total distracted time. We found significant differences between the control condition and the other conditions.}
        \Description{This bar chart with standard errors shows participant’s total distracted time during each session, comparing the mindless, alerting, and control condition. The time for the control condition is significantly longer than that for the mindless and alerting condition (**). The time for the mindless and alerting condition is not significant.}
        \label{fig:exp2-time}
    \end{minipage}
    \hfill
    \begin{minipage}[t]{0.33\linewidth}
        \centering
        \includegraphics[width=0.905\textwidth]{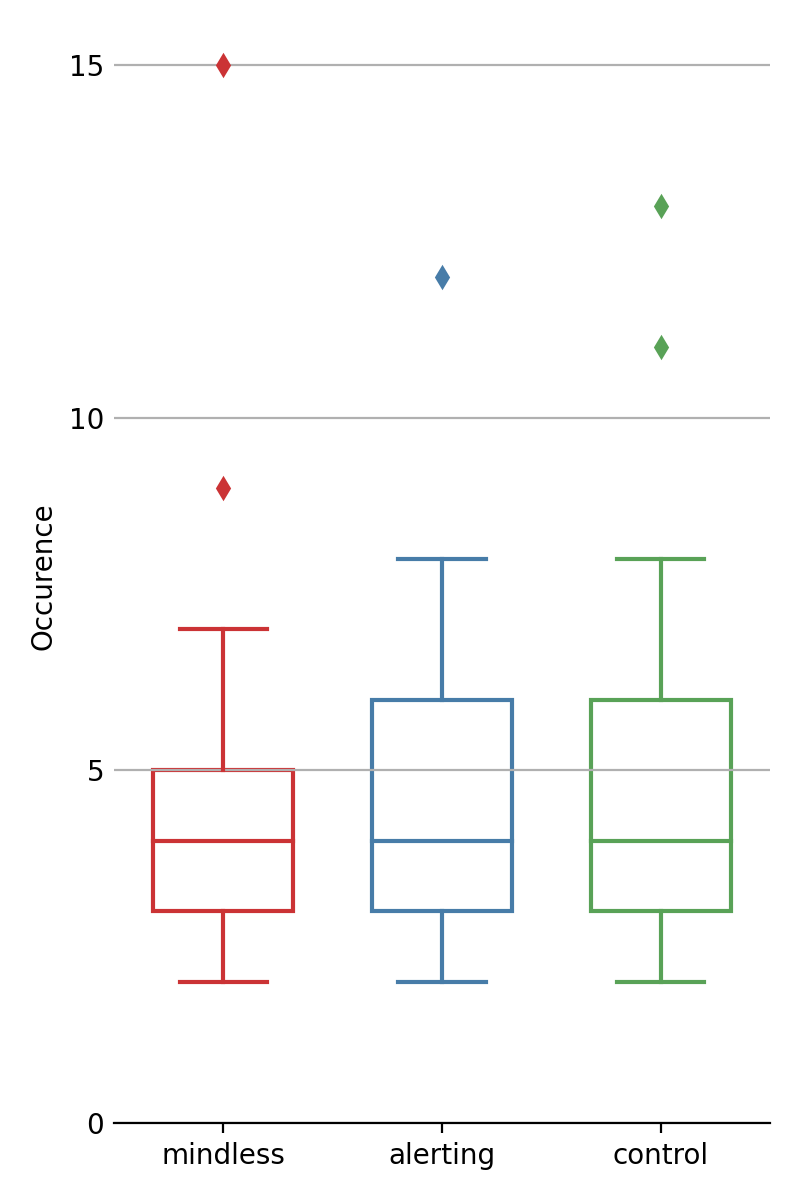}
        \caption{Comparison of how many times participants got distracted. We found no significant difference between the three conditions.}
        \Description{This bar chart with standard errors shows how many times participants got distracted during each session, comparing the mindless, alerting, and control condition. There is no significant difference between each condition.}
        \label{fig:exp2-occurence}
    \end{minipage}
    \hfill
    \begin{minipage}[t]{0.29\linewidth}
        \centering
        \includegraphics[width=0.76\textwidth]{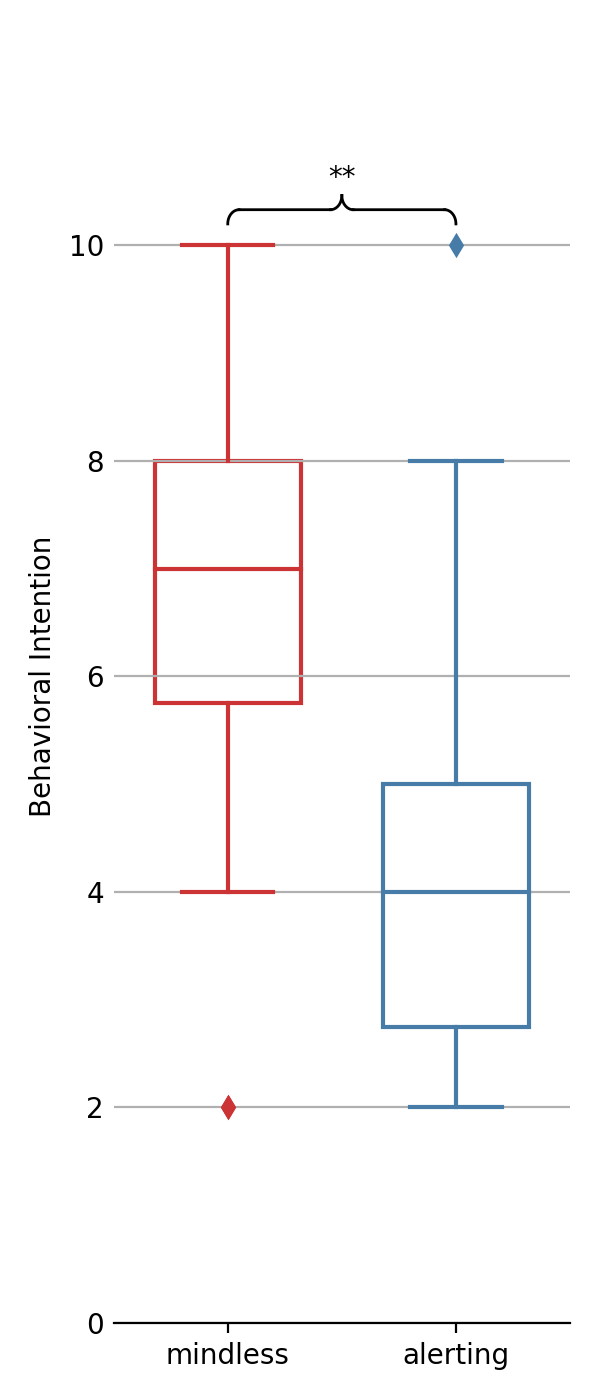}
        \caption{Comparison of participants' scores of the behavioral intention. We found a significant difference between the mindless and alerting conditions.}
        \Description{This bar chart with standard errors shows the scores of the behavioral intention, comparing the mindless and alerting condition. The score for the mindless condition is significantly higher than that for the alerting condition (**).}
        \label{fig:exp2-intention}
    \end{minipage}
\end{figure*}

Next, based on the human annotations, we calculated the total distracted time for each participant, as presented in \figref{exp2-time}.
We found a significant difference among the three conditions according to ANOVA ($F(2, 57) = 8.5773$, $\eta^2 = 0.2313$, $p = 0.0005$), and thus conducted a post-hoc test.
As a result, the control condition showed significant differences against the mindless and alerting conditions (Cohen's $d = 1.1795$, $p = 0.0013$ and Cohen's $d = 1.0828$, $p = 0.0032$, respectively).
On the other hand, we found no significant difference between the mindless and alerting conditions.

From this result, it was confirmed that Mindless Attractor is an effective means to refocus users' attention even when combined with a machine learning-based sensing module, as the mindless condition significantly reduced the total distracted time than the control condition.
In addition, it is notable that Mindless Attractor would work effectively as well as the conventional alerting approaches since the mindless and alerting conditions showed similar distracted times.

We also examined how many times the participants got distracted because it was possible that our interventions increased the frequency even though the total distracted time was reduced.
As shown in \figref{exp2-occurence}, we did not find significant differences among the three conditions ($F(2, 57) = 0.1796$, $\eta^2 = 0.0062$, $p = 0.8360$).
It can be explained as follows: the participants were almost equally likely to lose focus in all the three conditions; but, if there was an intervention, they often refocused their attention to the video earlier, as confirmed in the first experiment; as a result, their distraction time in the mindless and alerting conditions was significantly reduced than the control conditions.
From these results, we conclude that H2 was supported in terms of the effectiveness of Mindless Attractor.

\subsubsection{Behavioral Intention}

Lastly, we compared participants' scores of the behavioral intention between the mindless and alerting conditions.
As presented in \figref{exp2-intention}, we found a significant difference (Cohen's $d = 0.7025$, $p = 0.0054$) according to the paired $t$-test.
That is, compared to the alerting approach, the participants showed their stronger intentions to use the implemented system when it is combined with the mindless approach.
This result supports that Mindless Attractor is much preferred by users than the alerting approach, as we hypothesized as H2.

\subsubsection{Comments}

The above results coincided with H2; that is, Mindless Attractor helps participants refocus their attention and it is favored over a conventional alerting approach.
In addition, the comments obtained at the end of the experiment corroborated H2, especially in regard to the unacceptability of the alerting approach.
\begin{quote}
    I felt like the beep sound made me lose focus. It was frustrating, especially when I was concentrating. (P9)
\end{quote}
\begin{quote}
    The beep felt like noise because it overlapped the speech though I wanted to listen to what was being said. As a result, my concentration was more disrupted than the case that I had not used the system. (P12)
\end{quote}
\begin{quote}
    I thought the one with the beep sound might be a good signal until halfway through, but then it came to ring repeatedly even though I was concentrating. As a result, I stopped caring about the sound. (P2)
\end{quote}
These comments confirmed our anticipation; that is, explicitly alerting users based on false-positive detection makes them distracted and frustrated, which can lead them to ignore the intervention.
In addition, one participant suggested that such negative effects can be caused even when the intervention was activated by accurate detection:
\begin{quote}
    I was disgusted by the alarm, which rang when I was using my smartphone for googling a word I never heard. (P8)
\end{quote}

In contrast, the mindless condition was totally favored, as follows:
\begin{quote}
    In the part [of the mindless condition], I felt like I was able to focus on the lecture relatively well. (P12)
\end{quote}
\begin{quote}
    I did not notice much of a change in the audio, but when I compare the three parts, I seemed to be able to maintain my concentration the most. I think having such a system that brings back my attention without making a big deal will help me stay focused in usual situations. (P3)
\end{quote}
\begin{quote}
    When the pitch of the speech became higher, I paid attention to the video as I felt strange a little. It did not provide a sense of being angry, compared to the beep alarm. (P11)
\end{quote}
These comments corresponded to the comparison of the scores of behavioral intention (\figref{exp2-intention}).

Furthermore, 17 of 20 participants agreed they often have trouble maintaining their attention and computationally solving it would be beneficial, like:
\begin{quote}
    I find it difficult to maintain my attention in such online situations because of the lack of eyes around. (P1)
\end{quote}
In addition, they suggested that the proposed approach can be used outside video-based learning situations.
\begin{quote}
    I thought it would be nice to be able to introduce a similar system in offline situations. I will appreciate it if some device such as a smartwatch helps me refocus when I am losing my attention from an important conversation. (P4)
\end{quote}
The obtained comments not only supported the effectiveness of Mindless Attractor through supporting H2 but also highlighted the further potential of the proposed approach.


\section{Discussion}
\label{sec:disc}

So far, by verifying H1 and H2, we have demonstrated that Mindless Attractor works effectively as a novel intervention approach to support users' participation during video-based learning.
In this section, we contemplate the findings of our study, envision future application scenarios, and discuss limitations and directions for future work to further pave the way for supporting users in video-based communication.

\subsection{Necessity of Mindless Intervention in Machine Learning-Based Systems}
\label{sec:disc-necessity}

The results of our second experiment supported H2: Participants favored the proposed mindless approach, while the alerting approach was not accepted.
Specifically, the obtained comments suggested that participants were annoyed by the alerts when they were triggered by false positives of the sensing module.
In other words, mistakenly intervening in an explicit manner while users are concentrated on the main task can unnecessarily consume their conscious awareness and eventually disrupt their experience.

Indeed, such failures in designing automated systems based on machine learning-based sensing modules have been pointed out in a recent guideline for human--AI interaction \cite{Amershi2019Guidelines}.
That guideline emphasized the importance of considering that such AI-infused systems may demonstrate unpredictable behaviors due to false positives and false negatives.
Consequently, it was suggested that an effective approach in designing AI-infused systems is to enable users to dismiss the undesired functions instantly.
In light of this, our proposed mindless approach can be a promising direction that follows this guideline, as it does not consume users' conscious awareness, letting them not mind the mistakenly triggered interventions without much cognitive workload.
Therefore, we believe that Mindless Attractor can support users as a novel intervention method integrated with machine learning-based systems in various cases, not limited to the presented case (i.e., video-based learning).

\subsection{Application Scenarios}

As mentioned in \secref{intro}, the importance of helping participants be attentive during video-based communication has been emphasized in various contexts.
In this regard, we believe that Mindless Attractor can be used effectively not only in video-based learning but also in other situations using video-based communication.
For example, it can be employed to help participants in video-based meetings be more attentive in the same manner as shown in this study.
Here, we note that a few studies have aimed to provide real-time feedback to participants in meetings \cite{DBLP:conf/iui/SchiavoCMSZ14,DBLP:journals/imwut/SamroseZWLNLAH17}.
For example, CoCo is a system designed to achieve balanced participation through feedback, such as showing a pie chart representing the participation ratio that can be estimated from speaking length and frequency \cite{DBLP:journals/imwut/SamroseZWLNLAH17}.
Similar to the discussion we had with regard to video-based learning, these techniques of providing explicit feedback require participants to be motivated to change their behavior, i.e., to be more attentive to the meetings based on the feedback.
Therefore, we can expect that Mindless Attractor will be a promising alternative approach in that it does not consume participants' conscious awareness during meetings, even when combined with machine learning-based sensing systems.

Furthermore, we envision a future where Mindless Attractor can be utilized in everyday interpersonal interactions.
If we can assume that wearing earphones in our daily life become more popular, it is possible to perturb the sound they hear to utilize Mindless Attractor.
For example, once the system detects that the user is failing to pay attention during a conversation based on their behavioral or physiological data, the envisioned system can intervene in a mindless manner by modifying the voice they hear.
Note that such demand for offline use was indeed observed in one participant's comment (P4) in our second experiment.

It is noteworthy that we verified the effectiveness of Mindless Attractor in the experiments in which users used it with prior consent.
This lets us imagine further practical applications utilizing Mindless Attractor as an opt-in function.
More specifically, it would allow users to selectively turn the system on and off on their own, according to their situations and motivations.
For example, if a user attends an important lecture or meeting and thinks that they need the assistance, they can actively allow themselves to be exposed to the mindless intervention by turning on the system.
In other words, our results, which showed that the mindless approach worked with opt-in consent, will pave the way for the user-centered exploitation of computational interventions with which users can augment their levels of attention.

\subsection{Limitations and Future Work}

Though our experiments have demonstrated that Mindless Attractor is a promising approach, there are some limitations.
Initially, further investigations involving a greater number of participants and diverse lecture content are desirable to generalize our results.
For example, if a lecture is so attractive that learners are not distracted from the video, the proposed approach would not be necessary, while at worst it would not be harmful, as its impact on cognitive load was not observed in \secref{exp1-results}.

Secondly, our approach and evaluations are based on the discussion of Mindless Computing proposed by Adams et al. \cite{DBLP:conf/huc/AdamsCJC15}, considering users whose motivation for obeying the intervention is not always assumed.
In fact, we have given some consideration to the experimental designs so that the participants would not become much motivated to the video, like allowing the use of smartphones.
Thus, we skipped the measurement of the participants' motivation in our studies.
However, this means that their results would not necessarily guarantee the universal effectiveness of the proposed method for users with any levels of motivation.
Thus, evaluating participants' motivation and exploring its correlation with the efficacy of Mindless Attractor can be a promising future work.

In addition, the accuracy of the machine learning-based sensing module in the second experiment can be improved using the latest techniques \cite{DBLP:conf/icmi/ThomasJ17,DBLP:journals/ejivp/ZaleteljK17,DBLP:conf/um/HuttMBKBD17,DBLP:conf/ACMse/VeliyathDAHM19}.
In this study, we used a na\"{i}ve approach based on head pose to investigate the effect of the proposed approach with false-positive detection.
Although our sensing approach achieved a certain level of accuracy, as discussed in \secref{exp2-results}, there is room to further sophisticate the algorithm.
It remains to be explored how users would feel if the alerting approach is combined with a much more accurate sensing module.
Nevertheless, we believe that our mindless approach can be an effective intervention because false positives will still remain.

In relation to this, it is noteworthy that recent works have proposed methods for drowsiness detection from human visual cues \cite{Ghoddoosian2019Realistic}.
Thus, it can be explored in future work whether Mindless Attractor can help participants who get sleepy during video-based learning, by integrating such a detection technique in the sensing module.
Examining the boundary of the effectiveness of the proposed approach in such a situation would inform us of further possible approaches, such as a hybrid of the mindless and alerting interventions.

We also acknowledge that refining the design of alerts can mitigate the negative impact suggested in the second experiment. 
While we used a simple beep as an alert, alternative methods to inform users in less annoying manners are possible.
In particular, Weiser and Brown conceptualized ``calm technology'' as a more acceptable communication channel from computers \cite{Weiser1996Designing, Weiser1997Coming}.
For example, alerting users with less explicit sounds (e.g., birds chirping) could be preferred to a simple beep sound.
In addition, if we ignore the requirement of using the auditory modality, showing a status lamp on display is an alternative to inform users that they are losing attention.
However, as Adams et al. pointed out, these techniques require users' conscious awareness (e.g., interpreting the status based on the lamp) to induce behavioral change \cite{DBLP:conf/huc/AdamsCJC15}, while mindless computing does not.
Therefore, Mindless Attractor can be differentiated from alerting approaches in that it can work without consuming users' conscious awareness, as suggested in the first experiment (see \secref{exp1-results}).
That said, it is desirable to explore sophisticated alerting approaches to draw further implications in comparison to our mindless approach.

At the same time, the design of the mindless intervention has also room for exploration.
Currently, as explained in \secref{mindless-design}, we decided to perturb the pitch or volume of the voice based on the nature of human speech communication.
Though we did not statistically examine the results due to the small number of perturbations activated for each participant, there were individual differences in terms of their effectiveness, which would imply the possibility of personalizing the intervention patterns.
Moreover, human brains are known to show a special response to a self-voice \cite{CONDE201540} or a familiar voice \cite{PMID:16819998}.
Thus, a possible intervention might involve computationally modifying a voice so as to be similar to a self-voice or familiar voice when learners are not paying attention.
This can be achieved through recent techniques for high-fidelity real-time voice conversion \cite{DBLP:conf/interspeech/TodaMB12,Arakawa2019}.

Looking toward production deployment, investigating whether the proposed approach that helps learners pay attention contributes to their learning performance could be a future study.
Considering that previous studies adopting explicit feedback to help learners pay attention have shown a positive impact on performance \cite{Baker2010Better,DBLP:conf/icmi/XiaoW16}, our mindless approach can be expected to have a positive effect.
This is because the mindless approach exhibited an effect on distracted time comparable to that of the alerting approach in \secref{exp2-results}, while showing no significant impact on the cognitive load in \secref{exp1-results}.
Examining the long-term effect of the proposed approach is also suggested for future work.
Though our design is based on the concept of Mindless Computing, which Adams et al. \cite{DBLP:conf/huc/AdamsCJC15} have described as having long-term effectiveness, it is difficult to deny, without further investigation, the possibility that users will become acclimated to the perturbations.
However, even in this case, the combination with voice conversion we mentioned above could be a remedy, as it enables as many patterns of interventions as the number of conversion targets.

\section{Conclusion}

We presented a novel intervention approach, Mindless Attractor, which helps users refocus their attention in a mindless manner.
The approach leverages the nature of human speech communication and perturbs the voice that users hear when they are losing their attention.
Our first experiment confirmed the effectiveness of Mindless Attractor in a video-based learning context by showing that it helped users refocus their attention without consuming their conscious awareness.
Moreover, through a comparison with a conventional alerting approach, our second experiment further supported the efficacy of our proposed mindless approach when integrated as an automated system with a machine learning-based sensing module.
Based on the results of the experiments, we discussed implications for utilizing mindless interventions, especially in tandem with machine learning-based sensing modules, and envisioned future application scenarios.
Our findings and discussion pave the way for developing novel mindless interventions that can be harnessed in human--AI symbiosis.

\begin{acks}
This work is partially supported by JST ACT-X, Grant Number JPMJAX200R, Japan.
Several components of the sensing module used in this study were offered by ACES Inc., Japan.
\end{acks}

\bibliographystyle{ACM-Reference-Format}
\bibliography{main}

\end{document}